\documentclass[11pt]{amsart}
\usepackage{amsbsy,amssymb,amscd,amsfonts,latexsym,amstext,delarray,
amsmath,graphicx} 

\newtheorem{thm}{Theorem}[section]

\newtheorem{lem}[thm]{Lemma}

\numberwithin{equation}{section}

\def\C{{\mathbb C}}

\def\N{{\mathbb N}}

\def\Z{{\mathbb Z}}
\def\R{{\mathbb R}}

\def\cD{{\mathcal D}}

\def\cS{{\mathcal S}}

\def\cU{{\mathcal U}}

\def\Cas{{\rm Cas}}

\def\Tr{{\rm Tr}}

\def\fg{{\mathfrak g}}

\title{Dirac spectrum and spectral action of SU(3)}
\author{Alan Lai, Kevin Teh}
\address{Department of Mathematics  \\
California Institute of Technology \\ 
Pasadena, CA 91125, USA}
\email{teh\@@caltech.edu}

\begin{document}
\maketitle

\begin{abstract}
We compute the Dirac spectrum of SU(3) for a one parameter family of Dirac operators, including the Levi-Civita, cubic, and trivial Dirac operators. We then proceed to compute the spectral action for the entire family.
\end{abstract}

\section{Dirac Laplacian of a Lie group}
In this section, we define the one-parameter family of Dirac operators of a Lie group equipped with a positive definite, symmetric bilinear form on its Lie algebra. This family was studied in \cite{agricola}. The definition of the Dirac Laplacian is the same as in \cite{SU2}, but we reproduce it here for convenience.

In this paper, we consider the one-parameter family of Dirac operators on $SU(3)$ corresponding to the one-parameter family of connections,

\[
\nabla _X^t := \nabla _X^0  + t [X,\cdot] ,
\]

with $\nabla_X^0$ being the trivial connection of the Lie group coming from the left-multiplication trivialization of the tangent bundle. $\nabla_X^1$ is the trivial connection of the right-multiplication trivialization.

The connections so defined are metric connections, with torsion given by
\[
T(X,Y) =(2t-1)[X,Y],
\]
and hence $\nabla_X^{1/2}$ is the Levi-Civita connection. The operator corresponding to $t=1/3$, is the cubic Dirac operator studied in \cite{cubic}. Let $\langle \cdot, \cdot \rangle$ be the metric on $G$ and let $\{X_k \}$ be an orthonormal basis with respect to the metric. The $\mathfrak{so}(\fg)$ connection $\nabla_X^t$ lifts to a metric $\mathfrak{spin}(\fg)$ connection $\widehat{\nabla_X^t}$ given by

\begin{equation}
\label{connect}
\widehat{\nabla_X^t} = \nabla_X^0 + t \frac{1}{4}\sum_{k,l}\langle X, [X_k, X_l] \rangle X_k X_l,
\end{equation}
see \cite{par}.

Let $\C l (\fg)$ be the Clifford algebra generated by $\fg$ and the relation 
\begin{equation}
XY +YX = -2\langle X, Y \rangle,
\end{equation}
and let $\cU (\fg)$ be the universal enveloping algebra. Then the Dirac operator, $\cD_t$ of the connection \ref{connect}, as an element of $\C l (\fg) \otimes \cU(\fg)$ is given by 

\begin{equation}
\cD_t = \sum_i X_i \otimes X_i + t H,
\end{equation}

where 

\begin{equation}
H = \frac{1}{4} \sum_{j,k,l}X_j X_k X_l \otimes \langle X_j, [X_k,X_l] \rangle.
\end{equation}

\section{Spectrum of Dirac Laplacian of $SU(3)$}
We claim that the spectrum of the Dirac Laplacian of $SU(3)$ is as follows.

\begin{thm}
\label{spec}
In each row of the table below, for each pair $(p,q)$ with $p,q$ in the set of parameter values displayed, the Dirac Laplacian $\cD_t^2$ of $SU(3)$ has an eigenvalue in the first column of the multiplicity listed in the center column.

Let 
\[
\lambda(u,v) = u^2 + v^2 +u v.
\]
Let 
\[
m(a,b) = \frac{(p+1)(q+1)(p+q+2)(p+1+a)(q+1+b)(p+q+2+a+b)}{4}
\]
We denote by $\N^{\geq a}$, the set $\{n \in \N : n \geq a\}$, and we take $\N$ to be the set of integers greater than or equal to zero.

\[
\begin{array}{|c|c|c|}
\hline
\mathrm{Eigenvalue} & \mathrm{Multiplicity} & \mathrm{Parameter~Values}\\
\hline
\lambda(p+3t,q+3t)& m(1,1) & p \in \N, q \in \N \\
\hline
\lambda(p+2-3t,q-1+6t)& m(-1,2) & p \in \N, q \in \N \\
\hline
\lambda(p+1,q+1) + 3(3t-1)(3t-2)& m(0,0) & p \in \N, q \in \N, (p,q)\neq (0,0) \\
\hline
\lambda(p+3-6t,q+3t)& m(-2,1) & p \in \N^{\geq 1}, q \in \N \\
\hline
\lambda(p-1+6t,q+2-3t)& m(2,-1) & p \in \N, q \in \N \\
\hline
\lambda(p+1,q+1) + 3(3t-1)(3t-2)& m(0,0) & p \in \N^{\geq 1}, q \in \N^{\geq 1} \\
\hline
\lambda(p+3t,q+3-6t)& m(1,-2) & p \in \N, q \in \N^{\geq 1} \\
\hline
\lambda(p+2-3t,q+2-3t) + 3(3t-1)(3t-2)& m(0,0) & p \in \N, q \in \N \\
\hline
\end{array} 
\]
\end{thm}

There is some flexibility in the set of parameter values. For instance in the second line, we could have used instead $p \in \N^{\geq 1}$ since for that line the multiplicity is zero whenever $p=0$.

\subsection{Spectrum for $t = 1/3$}
In this case, the expression of the spectrum becomes much simpler

\begin{thm}
\label{specThird}
The spectrum for the Dirac Laplacian $\cD_{1/3}^2$ of SU(3) is given in the following table.

\[
\begin{array}{|c|c|c|}
\hline
\mathrm{Eigenvalue} & \mathrm{Multiplicity} & \mathrm{Parameter~Values}\\
\hline
p^2 + q^2 + pq & 2 p^2 q^2(p+q)^2 & p \in \N, q \in \N \\
\hline
\end{array} 
\]
\end{thm}

\subsection{Derivation of the spectrum}
We compute the spectrum of the Dirac Laplacian of $SU(3)$ using a representation-theoretic approach. First, we recall the fact that the Dirac Laplacian of a Lie group may be represented in terms of the Casimir element.

\begin{thm} \cite{SU2} Theorem 2.3,
\label{Dt2}
Let $\pi$ be the natural homomorphism $\cU(\fg) \rightarrow \C l(\fg)$. Let $\rho$ be the Weyl vector, the half sum of all positive roots of $\fg$. Then,
$\cD _t ^2 = (\pi \otimes 1)T_t$ where $T_t \in \cU (\fg) \otimes \cU(\fg)$ is given by 
\begin{equation}
T_t = (1-3t)(1 \otimes \Cas + \Cas \otimes 1 - \Delta \Cas) + 1 \otimes \Cas + 9t^2 |\rho|^2.
\end{equation}
\end{thm}

Next we use the Peter-Weyl theorem to reduce the computation to understanding the irreducible representations of $SU(3)$. On irreducible representations, the action of the Casimir is well understood. For any irreducible representation $V_{\mu}$, of highest weight $\mu$, the Casimir acts as scalar multiplication by the scalar
\begin{equation}
\label{cas}
\pi_{\mu}(\mathrm{Cas})=(\mu + \rho, \mu + \rho) - (\rho,\rho),
\end{equation}
where $\rho$ is the Weyl vector, the half-sum of all positive simple roots. The pairing $(\cdot,\cdot)$ is the dual pairing, on the weight space of a nondegenerate symmetric bilinear form on the Cartan subalgebra of $\mathfrak{g}$. Such a pairing is necessarily a constant multiple of the Killing form, which for $SU(3)$ is given by
\begin{equation}
\label{killing}
\kappa(X,Y) = 6 \Tr(XY),
\end{equation} 
where the trace and multiplication are taken in the natural representation of $X,Y$ as $3\times3$ matrices. One may identify $\fg ^*$ with $\fg$ by identifying $\lambda \in \fg^*$ with the unique $X_\lambda$ such that $(X_\lambda,Y) = \lambda(Y)$, for all $Y \in \fg$. This is possible due to the nondegeneracy of the pairing on $\fg$. In this way, one defines the dual pairing on $\fg^*$. The particular pairing which occurs depends on the normalization of the Riemannian metric. More specifically, the pairing is related to the Casimir operator by the following relation (the Casimir in turn being determined by the metric on $\mathfrak{g}$). 

\begin{thm}  \cite{kostant} Given a symmetric nondegenerate bilinear form on $\mathfrak{g}$, the corresponding Casimir element, and bilinear form on weights are related by
\begin{equation}
\frac{1}{24}\Tr(\mathrm{ad} (\mathrm{Cas})) = (\rho,\rho).
\end{equation}
\end{thm}

Henceforth, we assume that the metric is normalized so that $(\rho,\rho)=3$. This leads to the simplest expressions for the spectrum.

Therefore in order to derive the spectrum of the Dirac Laplacian, we must first analyze the pairing of weights. We take for our basis of the Cartan subalgebra to be $\{H_1, H_2 \}$,
\begin{equation}
H_1 = 
\left(
\begin{array}{ccc}
1 & 0 & 0 \\
0 & -1 & 0 \\
0 & 0 & 0
\end{array}
\right),
~H_2 = 
\left(
\begin{array}{ccc}
0 & 0 & 0 \\
0 & 1 & 0 \\
0 & 0 & -1
\end{array}
\right).
\end{equation}
We identify weights concretely using the basis. I.e. for $\lambda \in \mathfrak{h}^*$, we identify $\lambda$ with $(\lambda(H_1),\lambda(H_2))$.
Then, the Weyl vector is given by $\rho = (1,1)$, and the weights $\lambda_1 = (1,0)$ and $\lambda_2 = (0,1)$ form an $\N$-basis of the highest weights of irreducible representations of $SU(3)$. The pairing of weights can be determined up to normalization, using duality, and the Killing form \ref{killing}, from which one deduces the relations
\begin{equation}
(\lambda_1,\lambda_1) = (\lambda_2,\lambda_2) = 2 (\lambda_1,\lambda_2).
\end{equation}

\begin{lem}
\label{caspq}
On the irreducible representation of highest weight $(p,q)$, $p,q \in \N$, the Casimir element acts by the scalar
\begin{equation}
\pi_{p,q}(\mathrm{Cas}) = (p^2 + q^2 + 3 p + 3 q + p q)(\lambda_1,\lambda_1).
\end{equation}
\end{lem}
For the normalization that we are considering, we have $(\lambda_1,\lambda_1)=1$.

We have listed the irreducible representations of $SU(3)$ as well as the action of the Casimir operator on them. To write down the spectrum of the Dirac Laplacian the only obstacle now is to understand the term $\Delta \Cas$ in Theorem \ref{Dt2}; i.e. we need to know the Clebsch-Gordan coefficients of the tensor products $V_{\rho} \otimes V_{(p,q)}$. These were computed in \cite{CG}. We recall the Clebsch-Gordan coefficients that we will need below.

\begin{lem}
\label{ClebGord}
The decomposition of $V_{\rho} \otimes V_{(p,q)}$ into irreducible representations is 
\begin{equation}
V_{\rho} \otimes V_{(p,q)} = \oplus_{\mu} V_{\mu},
\end{equation}
where the summands $V_{\mu}$ appearing in the direct sum are given by the following table:
\begin{equation}
\begin{array}{|c|c|}
\hline
\mathrm{Summand} & \mathrm{Parameter~Values}\\
\hline
V_{(p+1,q+1)} & p \in \N, q \in \N \\
\hline
V_{(p-1,q+2)} & p \in \N^{\geq 1}, q \in \N \\
\hline
V_{(p,q)} & p \in \N, q \in \N, (p,q)\neq (0,0) \\
\hline
V_{(p-2,q+1)} & p \in \N^{\geq 2}, q \in \N \\
\hline
V_{(p+2,q-1)} & p \in \N, q \in \N^{\geq 1} \\
\hline
V_{(p,q)} & p \in \N^{\geq 1}, q \in \N^{\geq 1} \\
\hline
V_{(p+1,q-2)} & p \in \N, q \in \N^{\geq 2} \\
\hline
V_{(p-1,q-1)} & p \in \N^{\geq 1}, q \in \N^{\geq 1} \\
\hline
\end{array} 
\end{equation}
\end{lem}

Each summand in the left column appears once if $(p,q)$ lies in the set of parameter values listed on the right column. For instance for $(p,q) =(1,1)$, the summand $V_{(p,q)} = V_{(1,1)}$ appears twice in the direct sum decomposition, since $V_{(p,q)}$ appears twice in the left column, and $(p,q)$ is in the set of parameter values in each of the two rows.

By combining Theorem \ref{Dt2}, Lemma \ref{caspq}, and Lemma \ref{ClebGord}, we obtain Theorem \ref{spec}. The multiplicities are obtained using the Weyl dimension formula
\begin{equation}
\dim V_{(p,q)} = \frac{1}{2}(p+1)(q+1)(p+q+2).
\end{equation}

When $t=1/3$, the formula for the Dirac Laplacian in Theorem \ref{Dt2} simplifies to 
\begin{equation}
\cD_{1/3}^2 = 1 \otimes \mathrm{Cas} + 3.
\end{equation}
Therefore, we no longer need to decompose any tensor products into irreducible components, and using just \ref{caspq}, one obtains Theorem \ref{specThird}.

\section{Spectral action of $SU(3)$}
In this section, we compute the spectral action, $\Tr f(\cD_t^2/\Lambda^2)$. In the case $t=1/3$, one may apply the Poisson summation formula as in \cite{uncanny} to quickly obtain the full asymptotic expansion for the spectral action. For general $t$ however, this approach no longer works. An expansion can however still be generated using a two variable generalization of the Euler-Maclaurin formula \cite{karshon}. However, this requires more work to produce, and produces the full expansion of the spectral action only if one assumes in addition that the test function $f$ has all derivatives equal to zero at the origin. Here, we compute the spectral action to order $\Lambda^0$.

\subsection{t = 1/3}

Let $f \in \cS (\R)$ be a Schwarz function. By Theorem \ref{specThird}, the spectral action of $SU(3)$, for $t = 1/3$ is given by
\begin{equation}
\Tr f(\cD_{1/3}^2/\Lambda^2) = \sum_{p=0}^{\infty} \sum_{q=0}^{\infty} 2 p^2 q^2(p+q)^2f(\frac{p^2+q^2+pq}{\Lambda^2}).
\end{equation}
In order to apply the Poisson summation formula, one needs to turn this sum into a sum over $\Z ^2$. For this purpose, one takes advantage of the fact that the expressions for the eigenvalues and multiplicities are both invariant under a set of transformations of $\N^2$ which together cover $\Z^2$. The linear transformations of $\N^2$ which together cover $\Z^2$ are
\begin{align*}
T_1 (p,q) &= (p,q), \\
T_2(p,q) &= (-p,p+q), \\
T_3(p,q) &= (-p-q,p), \\
T_4(p,q) &= (-p,-q), \\
T_5(p,q) &= (p,-p-q), \\
T_6(p,q) &= (p+q,-p). \\
\end{align*}
Each of the transformations is injective on $\N^2$. The union of the images is all of $\Z^2$. The six images of $\N ^2$ overlap on the sets $\{(p,q):p=0\}$ and $\{(p,q):q=0\}$. However, the multiplicity is equal to zero at these points, and so this overlap is of no consequence. Therefore, we may now write the spectral action as a sum over $\Z ^2$
\begin{equation}
\label{spactThird}
\Tr f(\cD_{1/3}^2/\Lambda^2) = \frac{1}{6}\sum_{p=-\infty}^{\infty} \sum_{q=-\infty}^{\infty} 2 p^2 q^2(p+q)^2f(\frac{p^2+q^2+pq}{\Lambda^2})\end{equation}

For a sufficiently regular function, the Poisson summation formula (in two variables) is
\begin{equation}
\label{psf}
\sum_{\Z^2} g(p,q) = \sum_{\Z^2}\widehat{g}(x,y).
\end{equation}

Applying Equation \ref{psf} to \ref{spactThird}, and applying the argument used in \cite{uncanny} we get the following result.

\begin{thm}
\label{spactThird}
Let $f \in \cS (\R)$ be a Schwarz function. For $t=1/3$, the spectral action of SU(3) is
\begin{equation}
\Tr f(\cD_{1/3}^2/\Lambda^2) = \frac{1}{3}\iint_{\R^2}x^2y^2(x+y)^2f(x^2+y^2+xy)dxdy ~\Lambda^8 + O(\Lambda^{-k}),
\end{equation}
for any integer $k$.
\end{thm}

\subsection{General $t$ and the Euler-Maclaurin formula}
The one-variable Euler-Maclaurin formula was used in \cite{CCRobWalk} to compute the spectral action of $SU(2)$ equipped with the Robertson-Walker metric. A two-variable Euler-Maclaurin formula may be applied here to compute the spectral action on $SU(3)$ for all values of $t$.

Let $m$ be a positive integer. Let $g$ be a function on $\R ^2$ with compact support. One instance of the two-variable Euler-Maclaurin formula is \cite{karshon}
\begin{equation}
\label{eulmac}
\sum_{p=0}^{\infty}\sum_{q=0}^{\infty}\mathrm{'} g(p,q) = L^{2k}(\frac{\partial}{\partial h_1})L^{2k}(\frac{\partial}{\partial h_2})  \int_{h_1}^{\infty}\int_{h_2}^{\infty}g(p,q)dp dq \vert_{h_1=0,h_2=0} +R_m^{st}(g).
\end{equation}

The notation $\sum\sum'$ indicates that terms of the form $g(0,q)$, $q\neq 0$, and $g(p,0)$, $p\neq0$ have a coefficient of 1/2, $g(0,0)$ has a coefficient of 1/4 and the rest of the terms are given the usual coefficient of 1. The operator $L^{2k}(S)$ is defined to be
\begin{equation}
\label{em}
L^{2k}(S) = 1 + \frac{1}{2!}b_2 S^2 + \ldots + \frac{1}{(2k)!}b_{2k} S^{2k},
\end{equation}
where $b_j$ is the $j$th Bernoulli number. The number $k$ is defined by $k = \lfloor m/2 \rfloor$
The remainder $R_m^{st}(g)$ is

\begin{equation}
\label{rem}
\begin{array}{l}
 R_m^{st}(g) =  \sum_{I \subsetneq \{1,2\}} (-1)^{(m-1)(2-|I|)} \times \\
\times \prod_{i \in I} L^{2k}(\frac{\partial}{\partial h_i})\int_{h_1}^{\infty} \int_{h_2}^{\infty} \prod_{i \notin I}P_m(x_i)\prod_{i\notin I} \left( \frac{\partial}{\partial x_i}\right)^m g(x_1,x_2)dx_1 dx_2 \vert_{h=0}.
\end{array}
\end{equation}
Equation \ref{eulmac} is proved in an elementary way in \cite{karshon}, by casting the one-variable Euler-Maclaurin formula in a suitable form, and then iterating it two times.

Using Theorem \ref{spec}, one may write the spectral action in terms of eight summations of the form

\[
\sum_{(p,q) \in \N ^2} g_i(p,q),
\] 
where
\[
g_i(p,q) = f\left( \frac{\lambda_i(p,q)}{\Lambda^2} \right)m_i(p,q), \quad i = 1,\ldots,8.
\]

The notations $\lambda_i(p,q)$ and $m_i(p,q)$ denote the eigenvalues and multiplicities of the spectrum in Theorem \ref{spec}.

One then applies the two-variable Euler-Maclaurin formula to each of the eight summations to replace the sums with integrals. Then to obtain an asymptotic expression in $\Lambda$, one controls the remainder, $R_m^{st}(g)$, to arbitrary order in $\Lambda$ by taking $m$ to be sufficiently large, and computes the big-$O$ behavior of the other integrals to arbitrary order in $\Lambda$ by applying the multivariate Taylor's theorem to a large enough degree. The terms in the Taylor expansions of the integrals yield the asymptotic expansion of the spectral action.

\subsection{Analysis of remainders}
Let us consider in detail the case $I = \{ \}$, of the remainder, \ref{rem}. The functions $P_m(x_i)$ are periodic, and hence bounded. Furthermore, they are independent of $\Lambda$. Therefore to study the big-$O$ behavior with respect to $\Lambda$ of the remainder,  \ref{rem}, we only need to estimate the integral
\begin{equation}
\iint \left| \frac{\partial ^m}{\partial p ^m} \frac{\partial ^m}{\partial q ^m} f(s \lambda(p,q)) m(p,q)\right|.
\end{equation}
 The integration happens over $(\R^+)^2 = [0,\infty)\times[0,\infty)$, and $m(p,q)$ is the multiplicity polynomial. The differentiated function is a sum of terms, whose general term is given by
\begin{equation}
C s^if^{(i)}(t \lambda)\lambda^{(a_1,b_1)(p,q)} \ldots \lambda^{(a_i,b_i)}(p,q)m^{(j,k)}(p,q),
\end{equation}
where $C$ is a combinatorial constant, $s = \Lambda ^{-2}$, and where $j$ and $k$ are less than or equal to $m$ and $0\leq i \leq 2m-j-k$ and
\[
\sum (a_i,b_i) = (m-j,m-k).
\]

Since $m$ is degree 4 in both $p$ and $q$, we know that $j \leq 4$ and $k \leq 4$. Since $\lambda$ is degree 2 in both $p$ and $q$ we know that each of the coefficients $a_k$, $b_k$ is less than or equal to 2. Therefore, one has the estimate
\begin{equation}
2 i \geq \sum a_i = m-j \geq m-4,
\end{equation}
and so
\begin{equation}
i \geq \frac{m-4}{2}.
\end{equation}
It is not too hard to see that 
\begin{equation}
\iint f^{(i)}(s \lambda)\lambda^{(a_1,b_1)(p,q)} \ldots \lambda^{(a_i,b_i)}(p,q)m^{(j,k)}(p,q)dpdq
\end{equation}
is uniformly bounded as $s$ approaches zero, and therefore we have that the integral has a big-$O$ behavior of $O(s^{\frac{m-4}{2}})$ as $s$ goes to zero.

The same argument gives the same estimate for the terms in the cases $I =\{1\}$ and $I=\{2\}$. Therefore we have shown
\begin{lem}
The remainder $R^{st}_m(g)$ behaves like $O(\Lambda^{-(m-4)})$ as $\Lambda$ approaches infinity.
\end{lem}

Since the sum in the Euler-Maclaurin formula, \ref{eulmac} gives only a partial weight to terms on the boundary, and since the functions $g_i(p,q)$, are at times nonzero on the boundary, $\{p=0\} \cup \{ q=0\}$ even when there are no eigenvalues there, we must compensate at the boundary in order to obtain an accurate expression for the spectral action.

In doing so, one considers sums of the form
\begin{align}
& \sum_{p=0}^{\infty}g_i(p,0), \\
& \sum_{q=0}^{\infty}g_i(0,q).
\end{align}
One treats these sums using the usual one-variable Euler-Maclaurin formula, which for a function, $h$, with compact support is
\begin{equation}
\sum_{p=0}^{\infty}h(p) = \int_0^{\infty}h(x) dx + \frac{1}{2}h(0) -\sum_{j=1}^{m} \frac{b_{2j}}{(2j)!}h^{(2j-1)}(0) + R_m(h),
\end{equation}
where the remainder is given by
\begin{equation}
\label{onerem}
R_m(h) = \int_0^{\infty} P_m(x) \left( \frac{\partial}{\partial x}\right)^{m} h(x)dx.
\end{equation}

The necessary estimate for the remainder \ref{onerem} is as follows.

\begin{lem}
$R_m(g(p,\cdot))$ and $R_m(g(\cdot,q))$ behave as $O(\Lambda^{-m+4})$ as $\Lambda$ approaches infinity. 
\end{lem}

To prove this, we observe that since the polynomial $P_m(x)$ is bounded and independent of $x$, we only need to estimate for instance
\[
\left| \int_0^{\infty} \left( \frac{\partial}{\partial x}\right)^{m} g_i(x,0)dx\right|.
\]
The function $g_i(x,0)$ is of the form
\begin{equation}
\label{diffand}
f(\frac{a x^2 + b x + c}{\Lambda ^2} + d)m(x,0),
\end{equation}
where $a,b,c$ are independent of $\Lambda$ and $x$, and $d$ is independent of $x$. The polynomial $m(x,0)$ is of degree 4 in $x$. Therefore, when one expands the derivative of \ref{diffand} using the product rule, the derivatives of $f(\frac{a x^2 + b x + c}{\Lambda ^2} + d)$ are all of order $j \geq m-4$. A simple inductive argument shows that the expansion of $(\partial / \partial x)^j f(\frac{a x^2 + b x + c}{\Lambda ^2} + d)$ under the chain rule the terms are all of the form
\[
\frac{1}{\Lambda^k} f^{(i)}(\frac{a x^2 + b x + c}{\Lambda ^2} + d) \alpha(x),
\]
where $k \geq j$, and $\alpha(x)$ is a polynomial. Finally we conclude the proof of the lemma by observing that 
\begin{equation}
\int_0^{\infty} \left( \frac{\partial}{ \partial x}\right)^j f(\frac{a x^2 + b x + c}{\Lambda ^2} + d)\alpha(x)dx
\end{equation}
is uniformly bounded as $\Lambda$ goes to infinity.

\subsection{Analysis of main terms}
With the remainders taken care of, one still needs to work out the big-$O$ behavior of the spectral action with respect to $\Lambda$ of the remaining terms coming from the two-variable and one-variable Euler-Maclaurin formulas. 

The calculation required is lengthy, but the technique is elementary. One changes variables to remove (most of) the $\Lambda$ dependence from the argument of the test function $f$. Then, one uses Taylor's theorem to remove the $\Lambda$ dependence from the limits of integration, and whatever $\Lambda$ dependence remains in the argument of $f$. In this way, one can obtain the big-$O$ behavior of the spectral action with respect to $\Lambda$ to any desired order. Here, we do the computation up to constant order in $\Lambda$. If one assumes that the test function $f$ has all derivatives equal to zero at the origin, then one obtains the asymptotic expansion to all orders in $\Lambda$.

To give a better idea of how the calculation proceeds, let us consider in detail a couple of terms coming from the Euler-Maclaurin formulas.

One term that appears upon application of the Euler-Maclaurin formula is
\begin{equation}
\int_{0}^{\infty}\int_0^{\infty}g_1(p,q)dpdq.
\end{equation} 
where
\begin{align*}
g_1(p,q) &= f(\frac{(p+3t)^2 + (q+3t)^2+(p+3t)(q+3t)}{\Lambda^2})\times \\
&\times \frac{1}{4}(p+1)(q+1)(p+q+2)(p+2)(q+2)(p+q+4)
\end{align*}

First, one performs the change of variables,
\[
x = \frac{p+3t}{\Lambda}, y = \frac{q+3t}{\Lambda},
\]
whereby one obtains
\begin{align*}
&\frac{1}{4}\int_{3t/\Lambda}^{\infty}\int_{3t/\Lambda}^{\infty} f(x^2 + y^2 +x y) (1-3t+x\Lambda)(2-3t+x\Lambda)\times \\
&\times(1-3t+y\Lambda)(2-3t+y\Lambda)(2-6t+x\Lambda+y\Lambda)(4-6t+x\Lambda+y\Lambda)\Lambda^2 dx dy
\end{align*}
Next, one does a Taylor expansion on the two lower limits of integration about 0. The first term in this Taylor series is obtained by setting the limits of integration to zero.

\begin{equation}
\label{zeroTerm}
\begin{array}{l}
\frac{1}{4}\int_{0}^{\infty}\int_{0}^{\infty} f(x^2 + y^2 +x y) (1-3t+x\Lambda)(2-3t+x\Lambda)\times\\
\times(1-3t+y\Lambda)(2-3t+y\Lambda)(2-6t+x\Lambda+y\Lambda)(4-6t+x\Lambda+y\Lambda)\Lambda^2 dx dy
\end{array}
\end{equation}

Remarkably, if one sums the analog of \ref{zeroTerm} for $g_1, \ldots, g_8$ one obtains the complete spectral action to constant order. All of the other terms which appear in the computation (of which there are many) cancel out, to constant order in $\Lambda$, in an intricate manner.

The end result of the calculation is the following.
\begin{thm}
\label{spact}
Let $f$ be a real-valued function on the real line with compact support. To constant order, the spectral action, $\Tr(f(\cD_t ^2 /\Lambda^2))$ of $SU(3)$ is equal to
\begin{align*}
&2\iint_{(\R^+)^2} f(x^2+y^2+xy)x^2y^2(x+y)^2dx dy ~\Lambda^8  \\
&+ 3(3t-1)(3t-2)\iint_{(\R^+)^2} f(x^2+y^2+xy)(x^4+2x^3y+3x^2y^2+2xy^3+y^4)dx dy~\Lambda^6 \\
&+ 9(3t-1)^2(3t-2)^2 \iint_{(\R^+)^2} f(x^2+y^2+xy)(x^2 + xy +y^2)dx dy~ \Lambda^4\\
&+ 6 (3t-1)^3(3t-2)^3\iint_{(\R^+)^2} f(x^2+y^2+xy)dx dy ~\Lambda^2\\
&+ O(\Lambda^{-1}).
\end{align*}
Here, the integrals are taken over the set $(\R^+)^2 = [0,\infty) \times [0,\infty)$. If $f$, in addition, has all derivatives equal to zero at the origin, then this expression gives the full asymptotic expansion of the spectral action.
\end{thm}
The linear transformations, $T_1 \ldots T_6$, are all unimodular, and the images of $(\R^+)^2$ cover $\R ^2$, up to a set of measure zero. Therefore, in the case of $t=1/3$, integrating over $\R^2$ multiplies the result by a factor of 6, and we see that Theorem \ref{spact} agrees with Theorem \ref{spactThird}

When computing the asymptotic expansion of the spectral action using the Euler-Maclaurin formula, as a result of the chain rule, the negative powers, $\Lambda^{-j}$ appear only with derivatives $f^{(k)}(0)$, $k \geq j$. This is why the terms of the asymptotic expansion vanish for negative powers of $\Lambda$, when the derivatives of $f$ vanish at zero.

\end{document}